\documentclass[12pt]{article}
\usepackage{group21}

\title{Algebraic Quantization on the Torus and Modular Invariance}

\author{\underline{J. Guerrero}\adr{1,2}, V. Aldaya\adr{1,3} \and\
            M. Calixto\adr{1}}

\address[1]{Instituto Carlos I de F\'\i sica Te\'orica y Computacional,
Universidad de Granada. Fuentenueva s/n, 18002 Granada. Spain.}

\address[2]{Dipartimento di Scienze Fisiche, Mostra d'Oltremare, Pad. 19,
80125 Napoli, Italy.}
\address[3]{IFIC, Centro Mixto Universidad de Valencia-CSIC, Burjasot 46100,
    Valencia, Spain.}

\def\Gt{$\widetilde{G}\;$}

\def\x{\vec{x}}

\def\w{\omega}

\def\z{\zeta}
\def\medio{\frac{1}{2}}

\def\idhbar{\frac{i}{\hbar}}
\def\nn{\nonumber}
\def\ni{\noindent}

\def\nn{\nonumber}
\def\ni{\noindent}

\def\z{\zeta}
\def\Gt{$\widetilde{G}\,\,$}

\def\Gt{$\widetilde{G}\;$}

 \def\z{\zeta}
 \def\w{\omega}
 \def\medio{\frac{1}{2}}
 \def\nn{\nonumber}
 \def\ni{\noindent}

\def\nn{\nonumber}
\def\ni{\noindent}

\def\be{\begin{equation}}
\def\ee{\end{equation}}
\def\bea{\begin{eqnarray}}
\def\eea{\end{eqnarray}}
\def\ba{\begin{array}}
\def\ea{\end{array}}

\def\z{\zeta}
\def\al{\alpha}

\def\Gt{$\widetilde{G}\,\,$}
\def\Gtm{\widetilde{G}\,}

\def\GP{G_{\cal P}}
\def\GH{G_{\cal H}}
\def\GC{G_C}
\def\HG{{\cal H}(\Gtm)}
\def\HGa{{\cal H}^\al(\Gtm)}

\def\k{\vec{k}}

\def\L{\vec{L}}

\def\x{\vec{x}}

\def\z{\zeta}
\def\medio{\frac{1}{2}}

\def\ni{\noindent}
\def\idhbar{\frac{i}{\hbar}}
\def\w{\omega}

\def\ra2{\sqrt{2}}

\def\bdm{\begin{displaymath}}
\def\edm{\end{displaymath}}

\def\be{\begin{equation}}
\def\ee{\end{equation}}

\def\bea{\begin{eqnarray}}
\def\eea{\end{eqnarray}}

\begin{document}
\maketitle


\section{Algebraic Quantization}

The aim of the Algebraic Quantization is the quantum description of a physical
system by means of the unitary and irreducible representations of its
symmetry group.
Two cases have to be
considered, corresponding to systems without constraints and to those with
constraints, respectively.


In the simplest case, the group \Gt of quantum symmetries will be a central
extension by $U(1)$ of the group $G$ of classical symmetries.
Then the starting point is a Lie group \Gt which is a principal bundle with
fiber $U(1)$ and base $G$. The group law has the generic form:
\be
g''=g'*g\,,\qquad
\z''=\z'\z \exp[i\xi(g',g)]\,,
\ee

\ni where $g'',g',g\in G$, $\z'',\z',\z\in U(1)$, and
$\xi: G\times G\rightarrow R$ is a 2-cocycle.
The representation is built by means of the left action of the elements of \Gt
on complex functions (wave functions) on the group manifold,
$\hat{g}\Psi(g')=\Psi(g*g')$,
verifying the $U(1)$-function condition (phase invariance of Quantum
Mechanics):
$\Psi(\z*g)=\z\Psi(g)\,,\,\,\,\,\forall g\in \Gtm\,,\,\,\forall \z\in U(1)$.

However, this representation is reducible (all right transformations commute
with it). To reduce it, we have to impose certain restrictions on the wave
functions in order to trivialize this right action. Some new concepts are
needed for this purpose. We call a subgroup $A\subset \Gtm$ horizontal if
$A\cap U(1)={\bf 1}_{\Gtm}$ (which implies that the restriction of $\xi$ to
$A$ is a coboundary, that is, $\xi(a_1,a_2)=\eta(a_1*a_2)-\eta(a_1)-\eta(a_2)$,
for some funtion $\eta$ on $A$).
Taking into account that the group commutator is
$[g',g]=g'*g*g'^{-1}*g^{-1}$,
we define the characteristic subgroup $\GC$ as the maximal horizontal subgroup
such that the commutator group
$[\GC,\Gtm]$ is also horizontal.
$\GC$ contains those transformations which do not possess dynamical
(symplectic) content (such as time evolution, rotations, gauge symmetries).

The following step is to introduce the concept of polarization subgroup
$\GP$, as a maximal horizontal subgroup which includes the characteristic
subgroup, $\GC\subset\GP$.
To reduce the representation we impose the polarization condition on the wave
funtions:
\be
\Psi(g*\GP)=\Psi(g)\, .
\ee

In this way, we obtain a unitary irreducible representation of the group
\Gt on polarized $U(1)$-function on the group by means of its left action.


The second case is a non-trivial generalization of the formalism consisting
in substituting the structure group $U(1)$ of phase invariance by a bigger
group $T$ (to account for ``gauge'' invariance, constraints, etc.,
see \cite{Comm1}).

With this generalization, the group \Gt becomes a principal fibre bundle with
structure group $T$. The group $T$ itself will be a (non-trivial, in
general) central extension by $U(1)$, and will be, in general, non-abelian.

Let's consider a UIR $D$ of $T$ on a complex vector space $E$. If $T$ is
non-abelian, its representations can have dimension greater than 1.
The Hilbert space $\HG$ is made out of those $E$-valued functions on the group
\Gt polarized (as in the case of structure group $U(1)$)
verifying the $T$-function condition from the left:
\be
\Psi(g_T*g)=D(g_T)\Psi(g)\,.
\ee

It must be stressed that the construction of the Hilbert space $\HG$ depends
on the particular choice of the UIR $D^\al$ of $T$, where $\al$ is an index
characterizing the representation. Therefore, we have {\it non-equivalent
quantizations} for each choice of non-equivalent representation $D^\al$ of
$T$ (in the sense of superselection sectors).

The quantum operators are defined as before. However, in general not all the
quantum operators preserve the Hilbert space $\HGa$, i.e. not all the (left)
transformations of \Gt are compatible with the $T$-function condition
(which is also imposed from the left).
Therefore, we define the subgroup of {\it good operators}, $\GH$,
as those preserving the Hilbert space $\HGa$. This subgroup can be
characterized by the condition:
\be
[\GH,T] \subset \hbox{Ker}\; D^{\alpha}(T)\,.
\ee

The rest of quantum operators, those not preserving the condition above, are
{\it bad operators}. Among them, there may be operators which are not so bad,
in the sense that they can be interpreted as quantization-changing operators,
taking the whole Hilbert space $\HGa$ to another
$\cal{H}^{\al'}(\widetilde{G})$, where
$\al$ and $\al'$ label non-equivalent representations.


Special care should be taken if the structure group possesses dynamical
(symplectic) content, i.e. the
2-cocycle $\xi$ is not a coboundary when restricted to $T$,
and we cannot impose, in general, the
whole group $T$ in the $T$-function condition (it would lead to
inconsistencies). He have to choose a polarization subgroup $T_p$ in $T$, and
impose the condition:
$\Psi(g_{T_B}*g)=D(g_{T_B})\Psi(g)$,
where $T_B=T_p \cup U(1)$ and $D$ is a representation of $T_B$.
Then we proceed in the same way as before, simply changing $T$ for $T_B$
everywhere.

%
%
%
%
%
%
%

\section{Quantization of the Heisenberg-Weyl group on the torus}

Now, as a direct application of the formalism introduced in the previous
section, let us consider the problem of the quantization of the torus as a
symplectic manifold. We can perform it considering \Gt as the
Heisenberg-Weyl (H-W) group, with group law:
\bea
\x{\:}'' & = &   \x{\:}'+ \x \nn \\
\z''   & = &   \z'\z \exp\{\idhbar m\w[(1+\lambda)x_1'x_2+\lambda x_1x_2']\}\,,
\eea

\ni and $T$ a fibre bundle with base
$\Gamma_{\L}\equiv \left\{e_{\k},\,\k\in Z\times Z \right\}$ and fibre
$U(1)$, where
$e_{\k}$ are translations of $\x$ by an amount of
$\L_{\k}\equiv (k_1 L_1,k_2 L_2)$ (therefore $\Gtm/T\sim T^2)$. $\lambda$
parametrizes different (equivalent) 2-cocycles.
The fibration of $T$ by $U(1)$ depends on the values of $m, \w, L_1$ and $L_2$,
and is, in general, non-trivial (see \cite{frachall} for a detailed
discussion). Two cases have to  be considered:
%
%


When $\frac{m\w L_1L_2}{2\pi\hbar}=n\in N$, the structure group is
$T=\Gamma_{\L}\times U(1)$, and the $T$-function condition reads
$\Psi(g_{T}*g)={\cal D}(g_T)\Psi(g)$, with ${\cal D}(e_{\k},\z)=\z D(e_{\k})$.
$D(e_{\k})$ is a representation of the group $\Gamma_{\L} \approx Z\times Z$.
 We shall restrict ourselves to the trivial representation
$D^0(e_{\k})=1$ for the time being, and the non-trivial ones will be obtained
later on.

The $T$-function condition is written as
$e^{\idhbar m\w[(1+\lambda)k_1L_1x_2 + \lambda k_2L_2x_1]}
    \Psi^{0}(\x+\L_{\k},\z)= \Psi^{0}(\x,\z)$.
This restriction on the wave functions has severe consecuences:\
(a) There exist only two possible polarizations\footnote{We are
considering only real polarizations. There exist also a complex polarization
leading to holomorphic wave functions.} leading to
$\Phi^{0}(x_1)\;$ and $\Phi^{0}(x_2)$ respectively, (b) The wave function is
distributional, with support on discrete,
equally spaced values, and (c) The dimension of the representations (and of
the Hilbert space) is $n$.

Explicitly, the allowed values for the coordinates are
$x_2=\frac{k}{n}L_2\,$ or $x_1=\frac{k}{n}L_1,\, k\in Z$, depending on the
polarization we choose.
The wave functions have the form
$\Phi^{0}(x_2)=\sum_{k\in Z}a_k \delta(x_2-\frac{k}{n}L_2)$,
with periodicity in the coefficients $a_k$, $a_k=a_{k+n}, \forall k\in Z$,
that allow to write it on a more compact form:
$\Phi^0(x_2)=\sum_{k=0}^{n-1} a_k \Lambda^0_k(x_2)$, where
\be
\Lambda^0_k(x_2)\equiv\sum_{k_2\in Z}\delta(x_2^{(k)}-k_2L_2)
             = \frac{1}{L_2}\sum_{q\in Z} e^{i2\pi q x_2^{(k)}/L_2}\,,
\ee

\ni with $x_2^{(k)}\equiv x_2-\frac{k}{n}L_2$.

The subgroup of good operators is:
$\GH = \left\{\z(\hat{\eta}_1)^{\frac{k_1}{n}}(\hat{\eta}_2)^{\frac{k_2}{n}},\;
\;
k_1,k_2\in Z,\,\,\z\in U(1) \right\}$,
with $\hat{\eta}_1\equiv e_{(1,0)}$ and $\hat{\eta}_2\equiv e_{(0,1)}$.

We can obtain the whole set of non-equivalent quantization acting with the bad
operators (those operators of \Gt that are not in $\GH$, see \cite{frachall}):
$\Phi^{\vec{\alpha}}(x_2)=\hat{\eta}_1^{\alpha_1}\hat{\eta}_2^{\alpha_2}
      \Phi^0(x_2)=\sum_{k=0}^{n-1} a_k\Lambda_k^{\vec{\alpha}}(x_2)$.

Bad operators,
in this simple case, can be interpreted as quantization-changing operators.
The range of inequivalent quantizations is given by
$\alpha_1\in [0,\frac{L_1}{n})$ and $\alpha_2\in [0,\frac{L_2}{n})$.


When\footnote{The irrational case requires techniques from noncommutative
geometry and will not be considered here.}
$\frac{m\w L_1L_2}{2\pi\hbar}=\frac{n}{r}\in Q$, the structure group
$T$ possesses dynamical (symplectic) content and we have to choose a
polarization subgroup $T_p$. Since $T$ has a characteristic subgroup
$\GC=\{ r\L_{\k}, \k\in Z\times Z\}$, $T_p=\GC\cup \{k\L_{\k_p}, k\in Z\}$,
where $\k_p=(1,0)$ or $(0,1)$.

The $T$-function condition is $\Psi(g_{T_B}*g)=
{\cal D}(g_{T_B})\Psi(g)$, where $T_B\equiv T_p\cup U(1)$. Let's consider
(for simplicity) the trivial representation
${\cal D}^0(g_{T_p},\z)=\z$ of $T_p$.
The two possible choices of $\k_p$ lead to non-equivalent representations,
of dimension $n$:

For $\k_p=(0,1)$ the wave functions $\Phi^0_\perp(x_2)$ are the same as
in the integer case. The difference is in the good operators,
$\GH^\perp =\left\{(\hat{\eta}_1)^{r\frac{k_1}{n}},\,
(\hat{\eta}_2)^{\frac{k_2}{n}},\;\;k_1,k_2=0,...,n-1 \right\}$. \\

For $\k_p=(1,0)$  the wave functions have support in the values
$x_2=k\frac{r}{n}L_2,\, k\in Z$, satisfy
$\Phi^0_\parallel(x_2+rk_2L_2)=\Phi^0_\parallel(x_2)$, and have the form
$\Phi^0_\parallel(x_2)=\sum_{k=0}^{n-1}a_k\Lambda^{r,0}_k(x_2)$,
where $\Lambda^{r,0}_k(x_2)\equiv
      \frac{1}{rL_2}\sum_{q\in Z} e^{i2\pi q x_2^{r,(k)}/(rL_2)}$, with
     $x_2^{r,(k)}\equiv x_2-\frac{k}{n}rL_2$. The good operators are:
$\GH^\parallel =\left\{(\hat{\eta}_1)^{\frac{k_1}{n}},\,
(\hat{\eta}_2)^{r\frac{k_2}{n}},\;\;
k_1,k_2=0,...,n-1 \right\}$.

The nontrivial representations of $T_B$ can be obtained as before, with the
action of the bad operators (see \cite{frachall}).
It should be stressed that, although
$\frac{m\w L_1L_2}{2\pi\hbar}=\frac{n}{r}$, the dimension of the quantum
representation is $n$, as in the integer case.
These representations can be interpreted in terms of a torus $r$ times bigger
in one direction, i.e. the area of the effective torus is $rL_1L_2$, and then
$\frac{m\w (r L_1L_2)}{2\pi\hbar}=n$. Therefore, the same results as in the
integer case apply, but changing
$L_2$ by $rL_2$ if $\k_p=(1,0)$ or $L_1$ by $rL_1$ if $\k_p=(0,1)$.
Another possibility is to interprete the wave functions as multi-valued
(r-valued) functions on the original torus, therefore building a vector
representation.

In conclusion, we can say that in the fractional case we have to substitute
the traditional $U(1)$-bundle over the torus by a vector bundle of rank $r$
and Chern class $n$. The operators of $T_B$ will act in a diagonal way but
those of $T$ that are not in $T_B$ will mix the different component of the
vector-valued function, building in this way a (r-dimensional) representation
of the whole group $T$.

In this particular case, and because the representations of $T$ are of finite
dimension (despite it has dynamical content), we could have
considered the whole group $T$ and its representations, without resorting
to its subgroup $T_B$ for imposing the constraints.

See \cite{frachall} for applications of this study to the Fractional Quantum
Hall Effect.

\subsection{Modular invariance}

We consider as starting group
\Gt the Schr\"odinger group (or $WSp(2,R)$ group). This group contains
$Sp(2,R)\approx SL(2,R)$ as a subgroup, representing the (linear) symplectic
transformations of the plane as a phase space.
We repeat the same procedure as before, with $T$ the same structure group, and
one obtains, essentially, the same wave functions, although new polarizations
are now allowed, related by modular transformations  (see \cite{SL(2Z)} for
the explicit computations).

The condition for good operators,
$[\GH,T] \subset \hbox{Ker}\; D^{\vec{\alpha}}(T)$, gives, besides the ones
obtained before, new good operators coming
from the $Sp(2,R)$ subgroup.
The result $SL(2,Z)\subset \GH$ would be expected (statement of
modular invariance of the quantum theory), but
the final result depends on the particular representation $D^{\vec{\alpha}}(T)$
chosen, and on the value of $\,\,\frac{m\w L_1L_2}{2\pi\hbar}=\frac{n}{r}$.
For the integer case
($r=1$) we obtain the result:
$SL(2,Z)\subset \GH \,\,\,\Leftrightarrow\,\,\, \vec{\alpha}=0\,
(\hbox{ i.e. trivial rep.), and $n$ even}$.
If  $n$  is  odd,  only  a subgroup  of modular transformations
preserves the Hilbert space ${\cal H}^{0}(\widetilde{G})$ (see \cite{SL(2Z)}
for details).
If a non-trivial representation $D^{\vec{\alpha}}(T)$ is chosen,
the result depends on the rational or irrational character of
$\frac{na_1}{L_1}$ and $\frac{na_2}{L_2}$. If any of them is irrational
then the whole $SL(2,Z)$ is bad. When they are rationals only a proper
subgroup of $SL(2,Z)$ survives as good operators.  Particularly interesting
is the case $\frac{na_1}{L_1}=\frac{na_2}{L_2}=\medio$, corresponding to
antiperiodic boundary conditions, and for which the same results as in the
trivial representation and $n$ odd is obtained, for all values of $n$.
For the
fractional case ($r\neq 1$), only a proper subgroup of $SL(2,Z)$ remains as
good operators.

These results agree with  those of Bos and Nair \cite{Bos-Nair} (although
they consider only the integer case),
but are in disagreement with the results of Iengo and Lechner \cite{Iengo},
who obtain no constraint on the values
of $n$ and the representation $D^{\vec{\alpha}}(T)$ corresponds to that of
antiperiodic wave functions, i.e.
$\frac{n\alpha_1}{rL_1}=\frac{n\alpha_2}{rL_2}=\frac{1}{2}$.

\section*{Acknowledgements}
This work is partially suported by the DGICYT. J. Guerrero thanks the
University of Granada for a postdoc grant.

\end{document}